# Two-Dimensional Ferromagnetism and Driven Ferroelectricity in van der Waals CuCrP$_2$S$_6$

Youfang Lai[1, †], Zhigang Song[1,2, †,*], Yi Wan[1], Mingzhu Xue[1], Changsheng Wang[1], Yu Ye[1,3], Lun Dai[1,3], Zhidong Zhang[4], Wenyun Yang[1,3], Honglin Du[1], Jinbo Yang[1,3,5]

1 State Key Laboratory for Mesoscopic Physics and School of Physics, Peking University, Beijing 100871, P. R. China

2 Department of Engineering, University of Cambridge, JJ Thomson Avenue, CB3 0FA Cambridge, U.K.

3 Collaborative Innovation Center of Quantum Matter, Beijing 100871, P. R. China

4 Institute of Metal Research, Chinese Academy of Science, Shenyang 110016, P. R. China

5 Beijing Key Laboratory for Magnetoelectric Materials and Devices，Beijing 100871, P. R. China

*Correspondence: Zhigang Song (szg@pku.edu.cn)

† Those author contribute equally to this work, and they should be viewed as first authors.

**Abstract**

Multiferroic materials are potential to be applied in novel magnetoelectric devices, for example, high-density non-volatile storage. Last decades, research on multiferroic materials was focused on three-dimensional (3D) materials. However, 3D materials suffer from the dangling bonds and quantum tunneling in the nano-scale thin films. Two-dimensional (2D) materials might provide an elegant solution to these problems, and thus are highly on demand. Using first-principles calculations, we predict ferromagnetism and driven ferroelectricity in the monolayer and even a few-layers of CuCrP$_2$S$_6$. Although the total energy of the ferroelectric phase of monolayer is higher than that of the antiferroelectric phase, the ferroelectric phases can be realized by applying a large electric field. Besides the degrees of freedoms in the common multiferroic materials, the valley degree of freedom is also polarized according to our calculations. The spins, electric dipoles and valleys are coupled with each other as shown in the computational results. In experiment, we observe the out-of-plane ferroelectricity in a few-layer CuCrP$_2$S$_6$ (approximately 13 nm thick) at room temperature. 2D ferromagnetism of few-layers is inferred from magnetic hysteresis loops of the massively stacked nanosheets at 10 K. The experimental observations support our calculation very well. Our findings may provide a series of 2D materials for further device applications.

**Introduction**

Multiferroic materials are materials exhibiting two or more primary ferroic properties and are promising for potential applications in the non-volatile storage devices controlled by an external electric field. Previous researches are mainly focused on the three-dimensional (3D) materials, such as perovskites $BiFeO_3$.[1-5] However, as the demanded scale of devices becomes small down to only a few nanometers[6,7], these nano-devices made from thin films of 3D materials suffer from the dangling bonds and the leakage current due to the quantum tunneling, leading to a poor performance. It is noteworthy that the "intrinsic" two-dimensional (2D) materials, i.e. the van der Waals layers, may shed light to these problems and provide a platform for a new generation of the magnetic and electronic devices in the nano-scale because of their large dielectric constant and atomic smooth surfaces.[8] Recently, 2D ferroelectric polarization has been found in SnTe and $In_2Se_3$ monolayers in the theory or experiment.[9,10] On the other hand, 2D ferromagnetism is also demonstrated in theory and then observed in the $CrGeTe_3$ and $CrI_3$ layers.[11,12] Furthermore, some research groups proposed multiferroic material by doping or modifying some monolayers, such as black phosphorus and graphene.[13-17] All of these discoveries attract great interest in the field of condensed matters.[18-20] 2D materials with spontaneous ferromagnetism and ferroelectricity have rarely been reported, especially in the experiments.

In this work, by using first-principles calculations we predict that 2D $CuCrP_2S_6$ is type-I multiferroic candidate (or a ferromagnetically ferroelectric material), where the ferromagnetic and ferroelectric polarization stems from Cr and Cu elements, respectively.[21] The magnetoelectric coupling stems from the spin-orbit coupling. In experiments, ferroelectricity is observed in 13.3 nm-thick $CuCrP_2S_6$ using piezo force microscope (PFM) at room temperature. The ferromagnetism is observed from the hysteresis loop of the massively stacked nanosheets. The magnetoelectric coupling is suggested from the in-plane magnetization curves of a single $CuCrP_2S_6$ flake. The experimental observations are in good accordance with our first-principles predictions. We further note that the valley degree of freedom in the ferroelectric phase is also polarized. All the above results make it a promising candidate for future nano-scale device.

**Results and discussions**

    **Theoretical model**

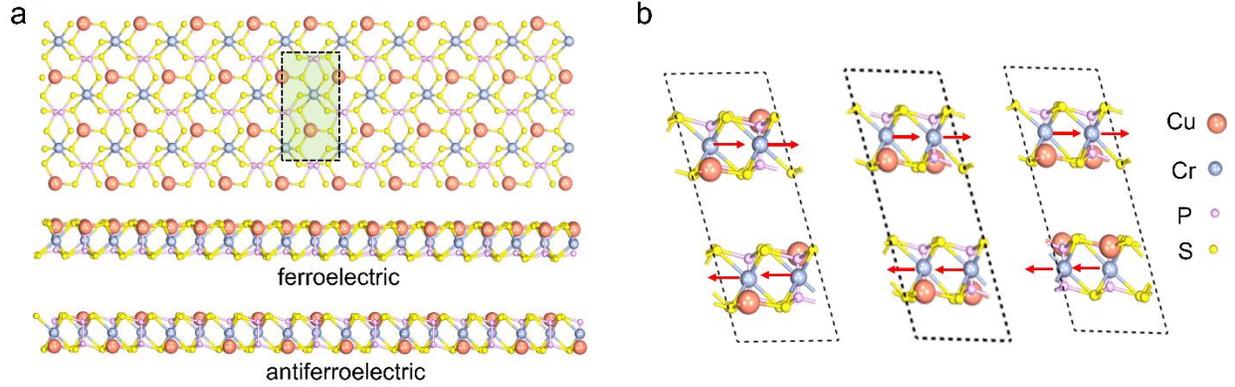

**Figure 1.** Optimized structures. (a) Top (up) and side (down) views of the $CuCrP_2S_6$ monolayer in ferroelectric and antiferroelectric phases. (b) Intralayer-antiferroelectric (left), ferroelectric (middle) and interlayer-antiferroelectric (right) bulks. Red arrows represent the directions of spins.

In this work, we take 2D $CuCrP_2S_6$ as an example to study low-symmetric $ABP_2X_6$ family[22] (A=Cu, Ag; B=Cr, V; X=S, Se)[23]. In $ABP_2X_6$ monolayers, six X atoms form the periodic rings, and A/B atoms center at the rings,[24] forming a large triangular periodic pattern. The next neighboring A or B atoms are bridged by a pair of phosphorus atoms. The optimized structure is shown in the Fig. 1(a). The equilibrium lattice parameters of the monolayer $CuCrP_2S_6$ are a=5.97 Å, b=10.36 Å, and β=104.1°, respectively. The bulk is stacked by a series of such layers (see Fig. 1(b)), and the interlayer distance is about 3.74 Å. Since the interlayer coupling is van der Waals interaction, the $CuCrP_2S_6$ bulk is possible to be exfoliated to monolayer and few-layer counterparts.

We firstly focus on the properties of $CuCrP_2S_6$ monolayer. In the optimized structure, Cu atoms have two distinct sites, leading to two possible phases.[8,25] One consists of Cu atoms lying on the same side of a monolayer (see Fig. 1(a)), resulting in an out-of-plane electric dipole. We label this phase as ferroelectric phase. The other consists of staggered Cu atoms siting in both sides of a monolayer. There are two opposite electric dipoles out of the *ab* plane in a unit cell, and the total electric dipole is zero. We label this phase as antiferroelectric phase. The displacement distance of the Cu cation between ferroelectric and antiferroelectric phases is about 2.65 Å and Born effective charge of Cu cation is approximately +1.2*e*. It should be noted that the electric dipole is approximately vertical to the *ab* plane. Our calculations of nudged elastic band (Fig. 2(a)) show that the antiferroelectric phase is the ground state, and ferroelectric phase is an excited state. Energy difference between the two phases is approximately 0.09 eV/ formula unit (f.u.). Nevertheless, it is noteworthy to mention the energy barrier between the two phases. On the one hand, the energy barrier from ferroelectric phase to antiferroelectric phase is approximately 0.11eV/(f.u.), which is relatively large for thermal perturbation (0.11eV/$k_B$~1300K). The ferroelectric phase won't deexcite very fast and thus can be observed once the ferroelectric phase is realized. On the other hand, the ferroelectric phase is easy to obtain by applying an external electric field due to low energy barrier (0.21 eV/ f.u.) in the nano-scale. Band structures with spin-orbital coupling (SOC) in Fig. 2(b) show that the band gaps of 2D ferroelectric and antiferroelectric $CuCrP_2S_6$ are 1.125 and 0.975 eV, respectively. The calculated absorption spectrums are shown in the Fig. S1

The calculated ground states of the $CuCrP_2S_6$ monolayer in both ferroelectric and antiferroelectric phases are ferromagnetic. The 3$d$ shell of Cr atom is not fully filled, leaving 3 unpaired $d$ electrons per atom, and thus the magnetic moment is 3 $\mu_B$/f.u. The ferromagnetic state is 13 meV/f.u. lower than the possible antiferromagnetic state where two neighboring spins are opposite. To determine the easy magnetization direction in ferroelectric and ferroelectric phase, we firstly calculate the energy difference between out-of-plane and in-plane spin configuration. The energy of the configuration that spin is aligned along $a$-axis is approximately 0.5 meV/f.u. lower than that of the spins vertical to the $ab$ plane, $E_a$-$E_v$ ≈ -0.5 meV/f.u. Such a small energy is calculated by linear extrapolation from the numerical results with artificial magnification of SOC strength (details can be seen in Fig. S2). In fact, the magnetocrystalline energy is about 1 meV/f.u. in most 3$d$ transition metals and their compounds[26], which is in good accordance with our results. The interesting thing is that the space group is P1 for ferroelectric monolayer, and the magnetocrystalline anisotropy energy is not necessary to respect to any symmetry. The direction of easy magnetization direction is related to the direction of the electric polarization, since the electric polarization can change the crystal field. For example, if the electric polarization is along positive $c$-axis, the energy of spins parallel to $a$-axis is about 1 meV/f.u. lower than that of spins antiparallel to $a$-axis, $E_a$-$E_{-a}$ ≈ -1.0 meV/f.u. When the electric polarization changes to negative c-axis, the energy difference will be $E_a$-$E_{-a}$ ≈ 1.0 meV/f.u. The large unidirectional magneto-crystalline anisotropy energy in the ferroelectric monolayers means the spins and electric dipoles are coupled in a sort of way.

Further investigation suggests that the electric dipoles, spins and momentum-retreated degrees of freedom, for example, valleys discussed in the next paragraph, are bridged by spin-orbit coupling $-\xi(\vec{d}\times\vec{P})\cdot\vec{S}$. The parameter $\xi$ is the coupling strength, and $\vec{d}$ is the electric dipole characterizing ferroelectricity. $\vec{P}$ is the quasi momentum, and $\vec{S}$ is the spin index, which is closely related to the intrinsic magnetization. The magneto-electric coupling is proportional to the momentum-resolved SOC in different bands by a constant. The integration over all bands and momentums is proportional to the total energy of SOC. SOC shows a nonzero contribution to the target system, and the integration of magneto-electric coupling is nonzero. Thus, the spins are locked with electric dipoles to some degree. When the electric dipole is reversed, the crystal field is changed, and the easy magnetization direction is reversed by an external electric field. Hence, the spins can flip in a monolayer without any an external magnetic field due to the unidirectional anisotropy. The illustration is given in Fig. 2(c).

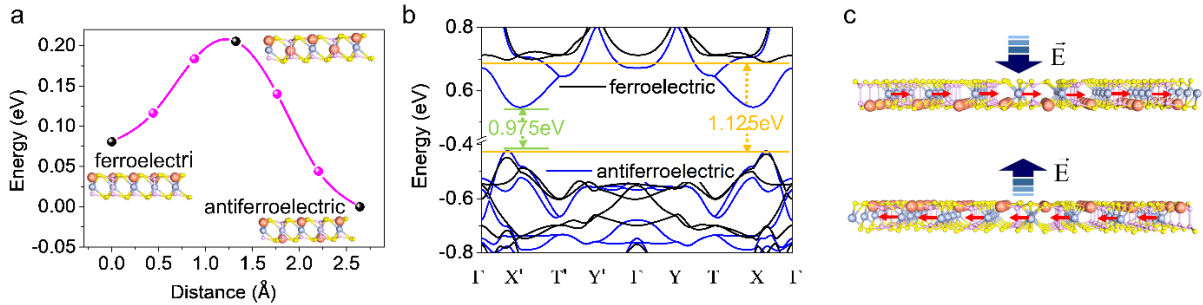

**Figure.2** Electronic structures of the $CuCrP_2S_6$ monolayer. (a) Phase transition path and corresponding barrier between ferroelectric and antiferroelectric phases. $d_{Cu}$ is the distance between two Cu sublayers. (b) Band structures with spin-orbit coupling for ferroelectric and antiferroelectric phases. (c) Illustration of locking between spin and dipole. Red arrows represent spins, and dash arrows represent external electric fields ($E$).

The spin-up and spin-down bands are well split owing to the ferromagnetic exchanging interaction. The bands near the Fermi level are dominated by the spin-up states in both ferroelectric and antiferroelectric phases. Both the conduction and valence bands are only projected on spin-up states. The corresponding spin-polarized band structures are plotted in Fig. S1(c), and the corresponding *k*-path of band strictures, high-symmetric points and Brillouin zone are given in Fig. S3. The band structures are featured by two massive Dirac cones, which constitutes a binary degree of freedom called valley degree of freedom.[27] Here, the two valleys located at (±0.67, 0, 0) will show opposite Berry curvature and orbital angular momentum. After SOC is included, the energy gaps at the two valleys remain the same in the antiferroelectric phase, and they are trivial. However, the direct band gaps at the two valleys in the ferroelectric monolayer are 1.125 eV and 1.145 eV, respectively. Hence the population of the two valleys will be different and the sample can show valley specified properties.

The different band gaps of different valleys in ferroelectric phase are analogues to the energy splitting in ferromagnetic phase, and the difference in the band gaps is called Zeeman-type valley splitting. More information and detail definition can be found in previous work.[22, 27-31] Here, the Zeeman-type valley splitting keeps a large value of 20 meV, which is a combined result of low crystal symmetry and SOC. The valley degree of freedom is strongly coupled to the electric and magnetic polarization according to the expression of the spin-dipole-valley coupling $-\xi(\vec{d}\times\vec{P})\cdot\vec{S}$, where $\vec{P}$ can characterize valley centers here. To obtain a general conclusion, we use the spins in the (111) direction to conduct the calculations (Fig. 3). It is interesting to note that the opposite electric polarization leads to opposite valley Zeeman splitting in the ferroelectric phase, if spins are fixed by an external magnetic field. Thus, the valley degree of freedom can fully be controlled by an external electric field, if spins are fixed. The Zeeman-type valley splitting can also be controlled by external magnetic fields. When the magnetization is reversed under the condition where the electric dipoles remain the same, the Zeeman-type valley splitting is reversed too.

It is worthy to stress that the roles of dipole, spin and valley are not equal. The energy to reverse valley, spin and dipole is different. Switching a dipole means that the crystal field is changed, and the energy needed to reverse a dipole is ~100 meV/f.u. Switching a spin involves in the energy of magnetocrystalline anisotropy, and the magnetocrystalline anisotropy energy is usually ~1 meV/f.u. In fact, valley is a slave, although many reports think valley is a degree of freedom. Thus, we can use electric field to control both magnetic moment and valley. It is hard to use magnetic field to control dipole only through the SOC, but the magnetic field can control the valley.

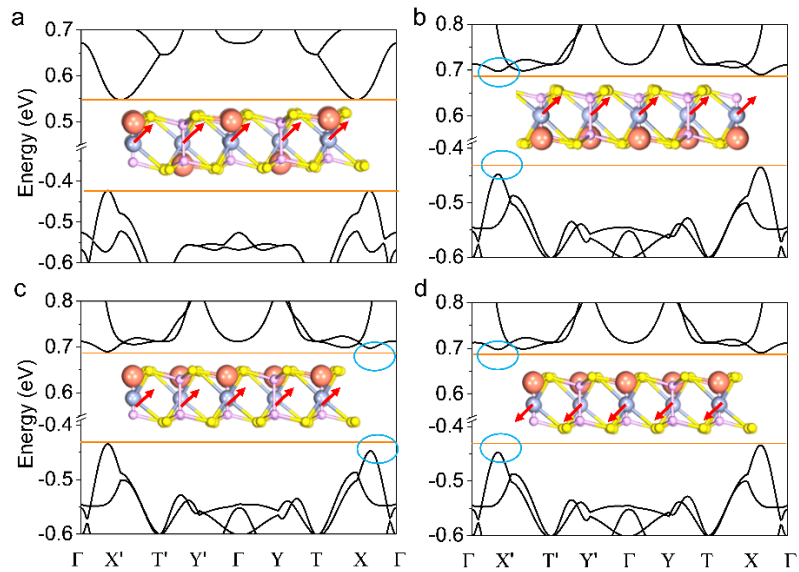

**Figure.3** Calculations of spin-valley-dipole locking. (a) Band structure with two trivial valleys in the antiferroelectric monolayer. (b-c) Band structures of ferroelectric monolayer with downward and upward electric polarization, respectively. (d) Band structure with two valleys in ferroelectric phase with upward electric polarization and opposite spins configuration compared to (c).

The bulk materials or nanosheets of few-layers can be viewed as stacked monolayers. The intralayer-antiferroelectric bulk is the most stable state according to calculations. The energy of ferroelectric bulk directly stacked by ferroelectric layers is 41 meV/f.u. lower than the energy of interlayer-antiferroelectric bulk. Thus, the ferroelectric stacking is easier to obtain than interlayer-antiferroelectric stacking. In few-layers, there are many kinds of possible stacking types for ferroelectric layers. The energy difference between different stacking types of ferroelectric layers is much smaller than the barrier from the ferroelectric phase to antiferroeletric phase in a monolayer. If we can polarize every layer by an external field, the ferroelectric stacking would be favored. Thus, stacking types can't depolarize the electric dipole. Many of them are ferroelectric or ferrielectric according to our calculations. Some examples of optimized structures of few-layers are given in Fig. S4. On the other hand, the nanosheets can be transited to ferroelectric phases by a large external field, even if the initial nanosheet is in intralayer-antiferroelectric phase.

Bulk (or 3D) $CuCrP_2S_6$ is antiferromagnetic, because the spins are oppositely aligned in neighboring layers no matter the bulk is ferroelectric or antiferroelectric. The calculated antiferromagnetic energy is about 1.5 meV/f.u. lower than that of the ferromagnetic state. The interlayer exchanging energy is much smaller than intralayer exchanging energy (13 meV/f.u.). Thus, intralayer magnetic ordering is more robust than interlayer magnetic ordering. The magnetization cannot be canceled in the nanosheets of odd layers. Even in nanosheets of even layers, ferromagnetic or ferrimagnetic phases possibly exist. According to our calculations, 4-layers can be ferromagnetic, and the magnetic configuration is shown in Fig. S4. As the number of layers increases, ferromagnetism can be neglected and the interlayer antiferromagnetism dominates. Similar results are observed in $VSe_2$ few-layers by experiment.[32] Overall, ferromagnetism is possibly observed in the few-layers of $CuCrP_2S_6$. According to the discussions above, due to the energy barrier between ferroelectric and antiferroelectric phase and some intermediate ferromagnetic states, ferromagnetic ferroelectricity is possible to exist in $CuCrP_2S_6$ few-layers.

The last but not the least, similar materials are predicted in the series of $ABP_2X_6$ mainly including $AgVP_2Se_6$, $CuCr_xIn_{(1-x)}P_2Se_6$ and $CuVP_2S_6$. Monolayer $AgVP_2Se_6$ has ferromagnetic ferroelectricity and a pair of valley pseudospins.[22] The electric polarization is in-plane, and the bulk is also ferromagnetic and ferroelectric. Monolayers of $CuCr_xIn_{(1-x)}P_2Se_6$ and $CuVP_2S_6$ are type-I multiferroic systems with out-of-plane electric polarization.

**Experimental results**

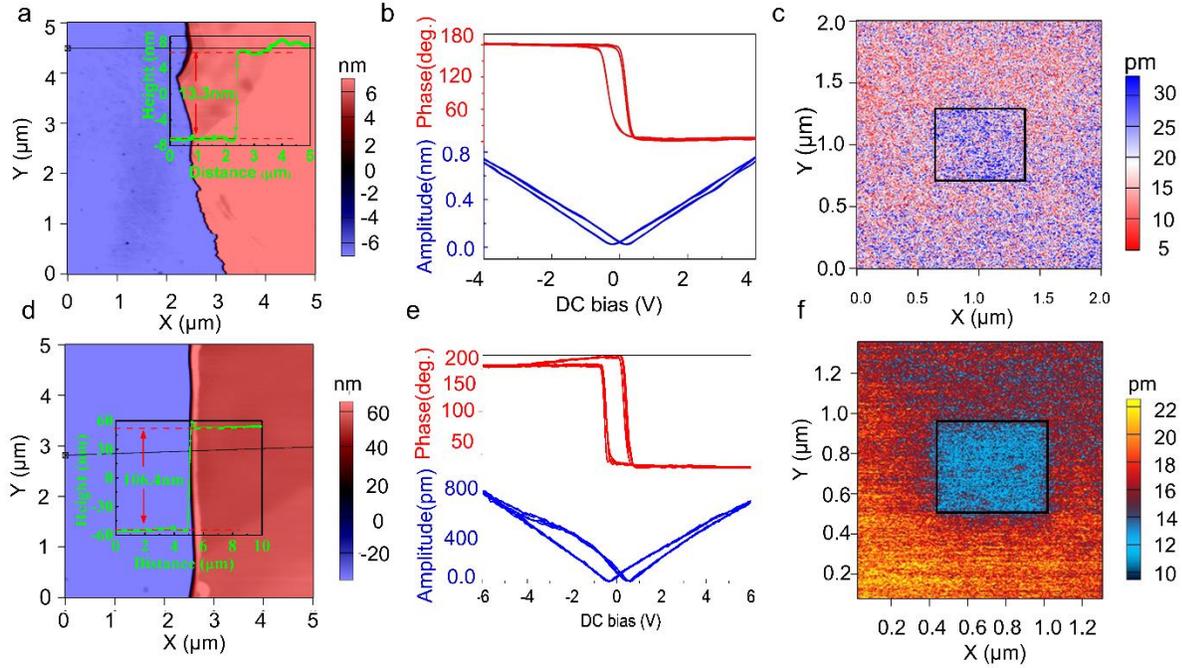

**Figure 4.** (a) Atomic force microscope image of $CuCrP_2S_6$ nanosheet with the thickness of approximately 13.3 nm. The inserted figure shows the height change along the black line. (b) Out-of-plane phase-voltage hysteresis loops (up) and amplitude-voltage butterfly loops (down) at room temperature. (c) Written pattern of 13.3 nm thick $CuCrP_2S_6$. (d-f) Characterization of thickness and ferroelectricity of 106 nm thick $CuCrP_2S_6$.

To support our theoretical proposal, the bulk and nanosheets of $CuCrP_2S_6$ were synthesized. The detailed preparation and characterization of the sample are given in the Method part. Fig. S5(b) shows the X-ray photoelectron spectroscopy (XPS) result of Cu:P3/2 and Cu:P1/2 peaks. All the peaks are accompanied with no sub peaks or satellites peaks, which indicates that Cu cation is of +1 valence.[33] The measured chemical valence is close to the calculated effective charge of +1.2 *e*. Nanosheets with thickness ranging from about 13.3 nm to several hundreds of nanometers are exfoliated from the bulk. The nanosheets are very stable and can survive for a few days without any protection, which is a great advantage for sample characterization. We use piezo force microscopy (PFM) to characterize the ferroelectricity of $CuCrP_2S_6$ nanosheets. Two nanosheets were selected to demonstrate a ferroelectricity measurement. The thickness of nanosheets measured by AFM is shown in Figs. 4(a, d), which is 13.3 nm and 106 nm thick, respectively. Figs. 4(b) shows the hysteresis loops of a selected thin $CuCrP_2S_6$ nanosheet with thickness of 13.3 nm. Saturated, closed and symmetric hysteresis loops of phase versus direct-current (DC) bias voltage and "butterfly" curves of polarized amplitude versus DC bias voltage are observed under resonance-enhanced mode at room temperature. The loops will not degrade after many measurement cycles. As shown in Fig. 4(b, e), they are consistent. The square hysteresis loops with a critical bias voltage of 0.32 V were obtained. Combined with the thickness of the nanosheets, the critical electric field is estimated to be $2.5\times10^7$ V/m. The polarization amplitude is of typical value as indicated in CuInP$_2$S$_6$,[34] and it shows the material is polarized. To further confirm the ferroelectricity, a 5 V drive voltage was used to write a rectangle pattern at a fast scanning rate for 20 cycles on the center area and then read the written signals on the whole area with 0.8 V driving voltage at a normal scanning rate. Because no corresponding damage in height image is observed after the writing processes, the obtained pattern as shown in Fig. 4(c) confirms that the ferroelectricity is sustainable and CuCrP$_2$S$_6$ is a 2D ferroelectric material. According to calculation, the ground state of the material is antiferroelectric in both bulk and monolayer materials. If we want to obtain the ferroelectric state, we need to apply electric field to excite the materials to ferroelectric state by driving the Cu ions. Since the electric field is positive (or negative) in the whole written area, a single domain tends to appear. This is the reason that no domain contrast is observed. Another nanosheet with a thickness of 106 nm shows a similar result as 13.3-nm-thick nanosheet. It is consistent with our analysis that the stacking of layers cannot cancel the electric polarization as long as the enough high electric field can be applied to drive it into the ferroelectric phase. The presence of ferroelectric polarization under an external electric field is in good accordance with our theoretical model. It should be emphasized that the electric polarization is observed at room temperature, and the polarization is out of the *ab* plane.

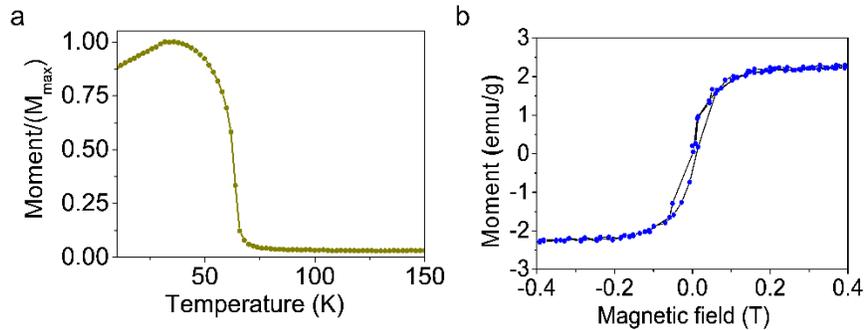

**Figure 5.** Characterization of magnetic properties in stacked nanosheets. (a) Thermal magnetic variation of massively stacked nanosheets of CuCrP$_2$S$_6$ with zero field cooling. Here M$_{max}$ is 0.0075 emu. (b) The magnetic hysteresis loops of massively stacked nanosheets of CuCrP$_2$S$_6$ at 10 K after subtracting the background. P (V) stands for parallel (vertical) direction to the *ab* plane.

Before measuring magnetic properties in nanosheets, we characterize the magnetism of a thick flake

(~ 1 μm thick). The result of the flake is similar to the previous bulk result despite of some differences.[35] The results are shown in Fig. S6, and more detailed discussions are given in the supplementary materials. No magnetic hysteresis loop is observed, and the magnetization versus the magnetic field is a linear function. But as we stated in the theoretical part, it can show intrinsic intra-layer ferromagnetism when thickness decreases. The measurement of micro magnetic properties of monolayer and few-layer nanosheets at very low temperature (below 31 K) is still a challenge due to the weak signal and the difficulty to measure in-plane magnetization. To enhance the signal, we used the grinding flakes, which we take it as a set of nanosheets with different thickness ranging from few to several hundred nanometers. The magnetization versus decreasing temperature, and hysteresis loop at 10 K are plotted in Fig. 5 (a) and (b), respectively. When the temperature is very low, the interlayer antiferromagnetic interaction would be favored, although it is relatively weaker than the intralayer ferromagnetism. Hence, the magnetization induced by magnetic field increases with increasing temperature below 31K, since the effective exchange interaction is weakened by thermal noise. It shows a cusp at Neel temperature (31K). The cusp at 31K is broader compared to bulk material, since the sample consists of nanosheets with different thicknesses and orientation. When the temperature crosses the Curie temperature of intralayer layer ferromagnetism (64K, which is the temperature point that has largest dM/dT), a sharp declination of total magnetization to almost zero is observed due to phase transition. The magnetic hysteresis loop at 10 K obviously shows ferromagnetic characteristics after subtracting the linear background, including antiferromagnetism, paramagnetism, diamagnetism. Owing to random orientation of the nanosheets, magnetoelectric coupling is not observed in the stacked nanosheets. Some possible evidences for magnetoelectric coupling in the thick oriented flake is discussed in the part of characterization of bulk in the supplementary materials. Besides, we also observe a weak hysteresis loop on the measurement of in-plane magnetization of exfoliated nanosheets versus magnetic field by magneto-optic Kerr effect (see Fig. S7) after subtracting the diamagnetic signal from the substrate. This also implies the in-plane ferromagnetism of few-layer $CuCrP_2S_6$, since the bulk is antiferromagnetic. The above results confirm the ferromagnetism in few-layer $CuCrP_2S_6$. It is worthy to mention that similar results are also obtained in 3D multiferroic $BiFeO_3$.[36-38] Although bulk $BiFeO_3$ is antiferromagnetic, weak ferromagnetism is observed in experiment.[36-38] The reversion of spin by electric field (or gates) is also observed.[39]

In this work, we present the first-principles analysis of a new 2D semiconductor material $CuCrP_2S_6$ and the experimental results on the ferroelectricity and ferromagnetism of $CuCrP_2S_6$ nanosheets. Its bulk is antiferroelectric and antiferromagnetic, but we observe ferroelectricity and ferromagnetism in its 2D nanosheets. The multiferroicity refers to the ferromagnetism and ferroelectricity of the monolayer. We further point out the polarized valley degree of this material can be a new kind of ferocity from the calculation results. In this work, we do not give experimental evidences for polarized valleys due to the difficulty in applying voltage vertical to the nanosheets when measuring the photoluminescence spectrum. Further efforts should be devoted to the synthesize and experimental characterization of monolayers at low temperature, since the multiferroicity of monolayer is much more robust and obvious than that of few-layers. Micro magneto-optical Kerr microscope and PFM cooled by liquid Helium are suitable setups for future characterization.

**Methods**

Calculation details: First-principles calculations are performed using the Vienna ab initio simulation package (VASP) in the basis set of the projector augmented waves. The cutoff energy is set as 400 eV. During the calculations of monolayers, a vacuum space larger than 15 Å is applied to decouple the periodic interaction. The Perdew-Burke-Ernzerh (PBE) functional is used to treat the exchange and correlation. The geometry and ions are fully relaxed to obtain a reasonable structure until all forces on every atom are smaller than 0.01 eV/Å. Dense k-point meshes of $11\times7\times1$ and $11\times7\times5$ are used for monolayers and bulks, respectively. The effective Hubbard U is tested, and DFT+U has no significant effect on the results of ferromagnetic ferroelectricity but has an influence on the valleys of the conductance band. DFT+U has no significant effect on the valleys of the valence band, and thus the valley-contrast transports are sound in case of *p*-doping.

Material preparation: The flakes of $CuCrP_2S_6$ are prepared by placing the corresponding elements of stoichiometric ratio in an $Al_2O_3$ crucible with a cover, which is then placed in an evacuated silica tubes. The tube is kept at 973K for two weeks.[40] Slightly additional P (5%) is used. The $CuCrP_2S_6$ flakes are selected from product carefully and confirmed with XRD and XPS. Magnetic properties are measured by vibrating sample magnetometer (VSM) option of physical properties measurement system (PPMS-9T, Quantum Design). The thermal magnetic data is collected at a rate of 3 K/min under 50 Oe. These flakes are exfoliated to nanosheets on the heavily doped Si substrate. The microstructures, thickness and electric polarization measurements are performed using modes of SEM, AFM and PFM of Asylum Research Cypher (Oxford Instruments) at room temperature. Here, the EDXS option of SEM is used to confirm the composition of the sample again. The phase-voltage hysteresis loops are measured under the resonance-enhanced PFM mode by applying an alternating-current (AC) electric field (0.1 Hz) superimposing on a DC triangle saw-tooth waveform. We choose different points with an enough distance to measure the hysteresis loops, and many cycles are conducted to exclude some accidental mistakes in each point. Magneto-optical Kerr effect measurement is conducted at 5K on the exfoliated nanosheets sitting on a heavily doped Si substrate. The optical configuration is set up to measure the in-plane magnetization.

## Supporting Information

The Supporting Information is available free of charge on the…

## Author Information

Corresponding Author: Zhigang Song: 0000-0002-8355-6498 (E-mail: szg@pku.edu.cn)

Y.F.L. and Z.S. contribute equally to this work.

The authors declare no competing financial interest.

## Acknowledgements:

Author Y.F.L. thanks the help of Mr. Liang Zha and Dr. Guangqiang Wang in sealing the silica tubes. This work is supported by the National Key Research and Development Program of China (No. 2017YFA206303), the Natural Science Foundation of China (Projects No. 11674005, 51731001, 11204110), National Materials Genome Project (2016YFB0700600), and the National Basic Research Program of China (No.2013CB932604).